\documentclass[a4paper,11pt]{article}
\RequirePackage{graphicx}
\usepackage{enumerate}
\usepackage{bm}
\usepackage{braket}
\usepackage{slashed}
\usepackage{comment}
\usepackage{color}
\title{Schwinger-Unruh-Hawking radiation on manifolds\thanks{Talk
presented at the Third Hermann Minkowski Meeting on the Foundations 
of Spacetime Physics}}
\author{Tomohiro Matsuda\thanks{Laboratory of Physics, Saitama Institute
of Technology, Fukaya, Saitama 369-0293, Japan}}
\date{September 12, 2023}
\begin{document}

\maketitle

\abstract{
The whole picture of gauge theory is described by manifolds, while the
field equation provides only a part (a section) of the manifold.
Just as a three-dimensional object is reconstructed from two planar
images, a monopole is constructed by combining two solutions.
The Schwinger and the Unruh effects and the Hawking radiation are the
production of particles out of the ``vacuum''.
If the '''vacuum'' on the manifold is properly defined, these phenomena
should be described as local phenomena.
However, calculations using the field equations have so far resulted in
unnatural extrapolations.
We present a method for properly defining the ``vacuum'' and
explain how to resolve the local particle production on manifolds.
By defining the Stokes phenomena on the manifold, the Schwinger effect
is naturally accompanied by the Unruh effect.
Also, unlike the conventional Unruh effect, calculations on manifolds do not
suffer from the entanglement between disconnected wedges.
}
\section{Introduction}
If particles are created from the ``vacuum'' in front of us, it is
natural to define that ``vacuum'' on the spot.
If this cannot be done, the theory should be considered somehow flawed.
The monopole solution is a simple illustration of the fact that the
field equations in field theory are somehow flawed.
They describe only a part (a section) of the theory, and the whole image
can be understood naturally by analyzing the manifold.
The Schwinger\cite{Schwinger:1951nm} and the Unruh
(Fulling-Davies-Unruh\cite{Fulling:1972md, Davies:1974th,Unruh:1976db})
effects  and Hawking radiation\cite{Hawking:1975vcx}, which deal with stationary 
particle production, have similar problems.
Looking at the field equations alone, the vacuum can only be defined in
the far asymptotic state.
On the other hand, in manifolds, tangent spaces can be defined at any point.
This tangent space is well known to be the definition of the natural
local vacuum in general relativity.\footnote{There are two definitions
of the local ``vacuum'' in general relativity.
One is the Lorentz frame in mathematics, which is a direct product
space, and the other is Einstein's local inertial system, which is
not a direct product.
The starting point for the field equations is the Lorenz frame (i.e,
mathematical covariant derivatives are defined in open coverings 
using local trivialization).
Although they are giving the same metric, we see that the difference
between the two has a crucial consequence in 
the case of Hawking radiation.}
However, when discussing Hawking radiation, such a local vacuum has never been
used for local particle creation in combination with the field equations.
There is something here that must be unraveled, and that is what
we have found in this work.

In this work\cite{Matsuda:2023mzr,Enomoto:2022mti}, stationary particle
production is 
naturally described as   
``the Stokes phenomena always appearing in the vacuum''.
Such a picture can only be obtained by considering particle production
and the vacuum definition on the manifold.

The topological properties of manifolds have been actively discussed,
but the characteristic properties of manifolds concerning particle
production have rarely been discussed.
In particular, as far as we know, the local ``vacuum'' has never been
defined on manifolds to explain local particle production. 

To get an overview and an idea, let us first have a short look at the
field equation for the Schwinger effect.
The simplest case is that the electric field is
spatially homogeneous and it is constant in the z-direction.
We introduce a complex scalar field $\phi$ of mass $m$ in the four-dimensional
Minkowski spacetime.
The action $S_0$ on the tangent space is 
\begin{eqnarray}
S_0&=&\int d^4x \left(\partial_\mu \phi \partial^\mu\phi^*-m^2
	       \phi\phi^*\right).
\end{eqnarray}
Introducing covariant derivatives, the partial derivatives are replaced
as
\begin{eqnarray}
\partial_\mu &\rightarrow& \nabla_\mu \equiv\partial_\mu+i q A_\mu,
\end{eqnarray}
where $A_\mu$ is a gauge field.
The ``vacuum'' is now defined  on the tangent space attached to $A_\mu=0$.
Assuming the limit where dynamics of the gauge field itself is negligible,
 the external gauge field is given by
\begin{eqnarray}
\label{eq-vecpot}
A^\mu&=&(0,0,0, -E (t-t_0)),
\end{eqnarray}
which explains the electric field strength $\vec{E}=(0,0,E)$.
Note that $t_0$ is an arbitrary parameter.
Although it is not often recognized, if the electric field is defined in
such a way as to restrict the degrees of freedom of the theory, then the
scope of this equation is also restricted to the vicinity of $t=t_0$,
since (strictly speaking) the open covering for
``this equation'' is only defined in the vicinity of $t=t_0$ .
In manifolds, such open coverings are used in bundles to cover the base 
space.

\begin{figure}[t]
\centering
\includegraphics[width=0.8\columnwidth]{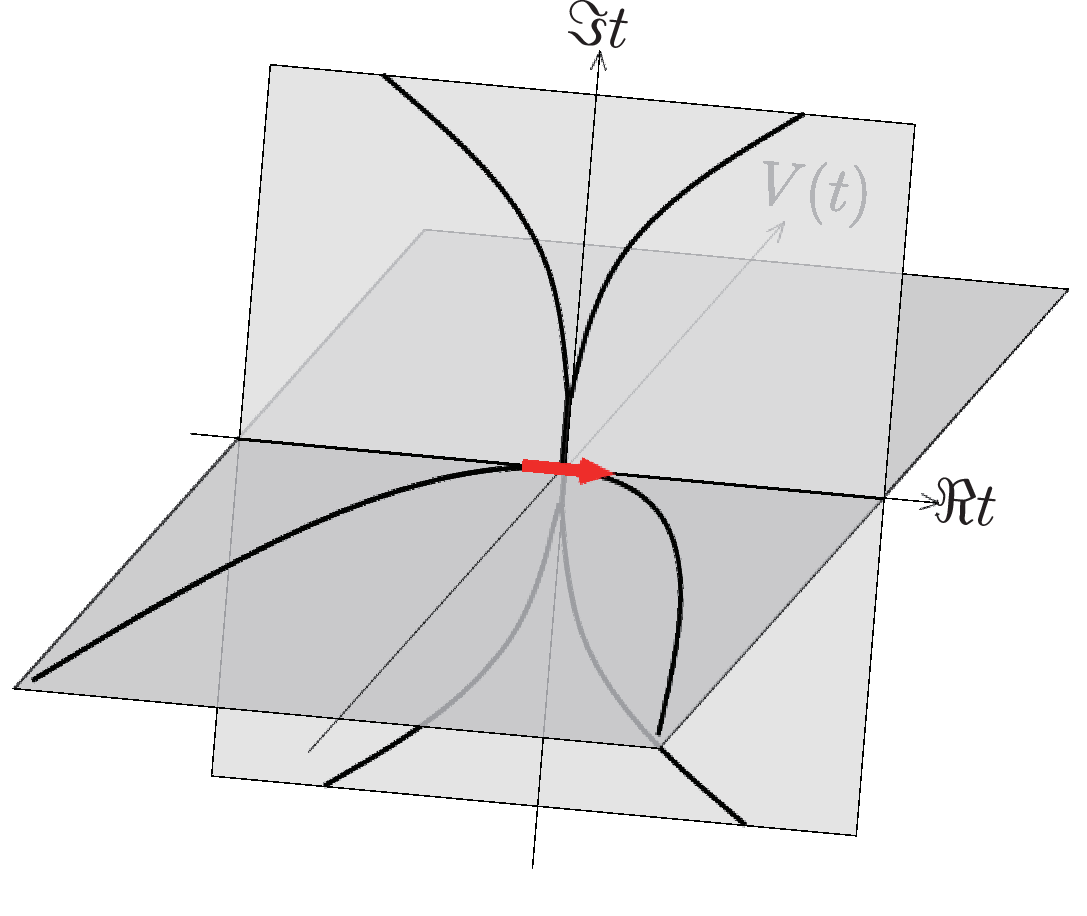}
 \caption{The potential $V(t)=-t^2$ and the Stokes lines (on the
 complex $t$ plane) are
 shown. Particle production occurs when the real $t$-axis crosses the
 Stokes line (indicated by the arrow at the origin).
For clarity, the Stokes line is written for $\omega^2_k(t) =1+t^2$, for which
two turning points appear on the imaginary $t$ axis. 
The original figure can be found in Ref.\cite{Matsuda:2023mzr}.}
\label{fig-1}
\end{figure}

The field equation for the scalar field $\phi$ after Fourier
transformation is
\begin{eqnarray}
\label{eq-field-Sch}
\ddot{\phi}_k+\omega_k^2(t)\phi_k&=&0,
\end{eqnarray}
where $\omega_k^2(t)=m^2+k_\perp^2+(k_z-qE(t-t_0))^2$.
We call $V(t)\equiv-\omega_k^2(t)$ a ``potential''.
Typical Stokes lines are shown in Fig.\ref{fig-1}.
Let us first think about $k_z=0$.
Since the Stokes line appears in the defined vacuum on the tangent
space, it can be seen that the vacuum solutions are mixed there.
The serious problem is that it is not apparent from the field equation alone
 why this mixing is occurring ``all the time''.
Also, it should be hard to recognize the definition of the vacuum only
by using the field equation. 
In order to understand the situation, we consider manifold for 
the field equation and the Stokes phenomena.

In the case of the Schwinger effect, the Stokes phenomenon can be seen
directly from the field equation, as we have described above.
Therefore, the problem can be solved when the equation is properly
considered on the manifold.
On the other hand, for the Unruh effect and Hawking radiation, 
the Stokes phenomena of particle production are not directly deducible
from the field equation. 
The answers to these questions are the subject of this
study\cite{Matsuda:2023mzr, Enomoto:2022mti}.

\section{How to define the Schwinger effect on the manifold}
Again, we consider Eq.(\ref{eq-field-Sch}) for the complex scalar field;
\begin{eqnarray}
\ddot{\phi}_k+\omega_k^2(t)\phi_k&=&0,
\end{eqnarray}
where $\omega_k^2(t)=m^2+k_\perp^2+(k_z-qE(t-t_0))^2$.

To describe the Bogoliubov transformation of the particle creation,
 we expand
$\phi_k(t)$ using the solutions $\psi^\pm_k(t)$ as 
\begin{eqnarray}
\label{eq-WKB}
\phi_k(t)&=& \alpha_k \psi^-_k(t)
+\beta_k\psi^+_k(t).
\end{eqnarray}
Then the transformation matrix is 
\begin{eqnarray}
\left(
\begin{array}{c}
\alpha_k^R\\
\beta_k^R
\end{array}
\right)
&=&
\left(
\begin{array}{cc}
\sqrt{1+e^{-2\pi \kappa}}e^{i\theta_1} & ie^{-\pi\kappa+i\theta_2}\\
-ie^{-\pi\kappa-i\theta_2} &\sqrt{1+e^{-2\pi \kappa}}e^{-i\theta_1} 
\end{array}
\right)
\left(
\begin{array}{c}
\alpha_k^L\\
\beta_k^L
\end{array}
\right),\nonumber\\
\end{eqnarray}
where the indices L and R are for $t<t_0$ and $t>t_0$, respectively. 
If one uses the exact WKB\cite{Enomoto:2020xlf, Enomoto:2021hfv,
Virtual:2015HKT}, the constant $\kappa$ 
can be calculated from the integral connecting the two 
turning points $t_*^\pm$ appearing on the imaginary axis.
The exact calculation gives 
$\kappa=\frac{m^2+k_\perp^2}{2E}$\cite{Kofman:1997yn, Enomoto:2020xlf,Enomoto:2021hfv}.
Here, all the phase parameters are included in $\theta_{1,2}(k)$.
It is already known\cite{Haro:2010mx} that Schwinger's original result
(the vacuum decay rate) can be calculated from the above result by
adding up all possibilities. 

\begin{figure}[h]
\centering
\includegraphics[width=0.8\columnwidth]{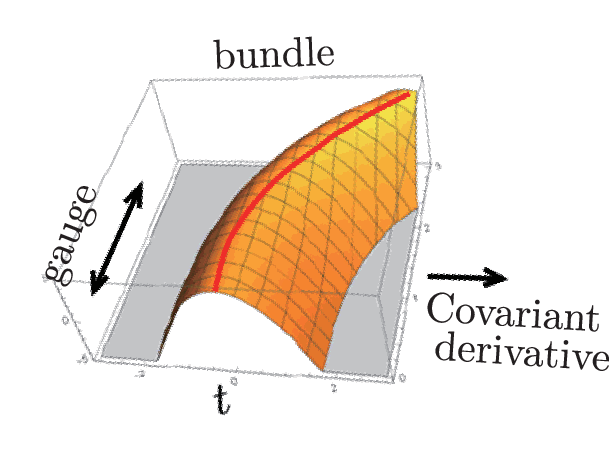}
 \caption{The field equation of the Schwinger effect is illustrated on
 the manifold. 
The Stokes phenomenon occurs in the ``vacuum'' (the tangent space)
 attached to the red line. 
The original figure can be found in Ref.\cite{Matsuda:2023mzr}.}
\label{fig-2}
\end{figure}

The obvious problem is that while the above equation describes particle
production at a specific time $t_0$, it should be arbitrary
because of the gauge degrees of freedom.
Therefore, we need to consider a specific way to incorporate the
degrees of freedom of the gauge into the equations of motion.
Our main claim is that this can be done on the manifold.
We have illustrated the situation in Fig.\ref{fig-2}.
We consider the equation with $\bf{\it k}$ set to 0 (or $\bf{\it k}^2\ll
\omega^2$), since the local
``vacuum'' is also defined for the rest frame of the particle.

\begin{figure}[t]
\centering
\includegraphics[width=1.0\columnwidth]{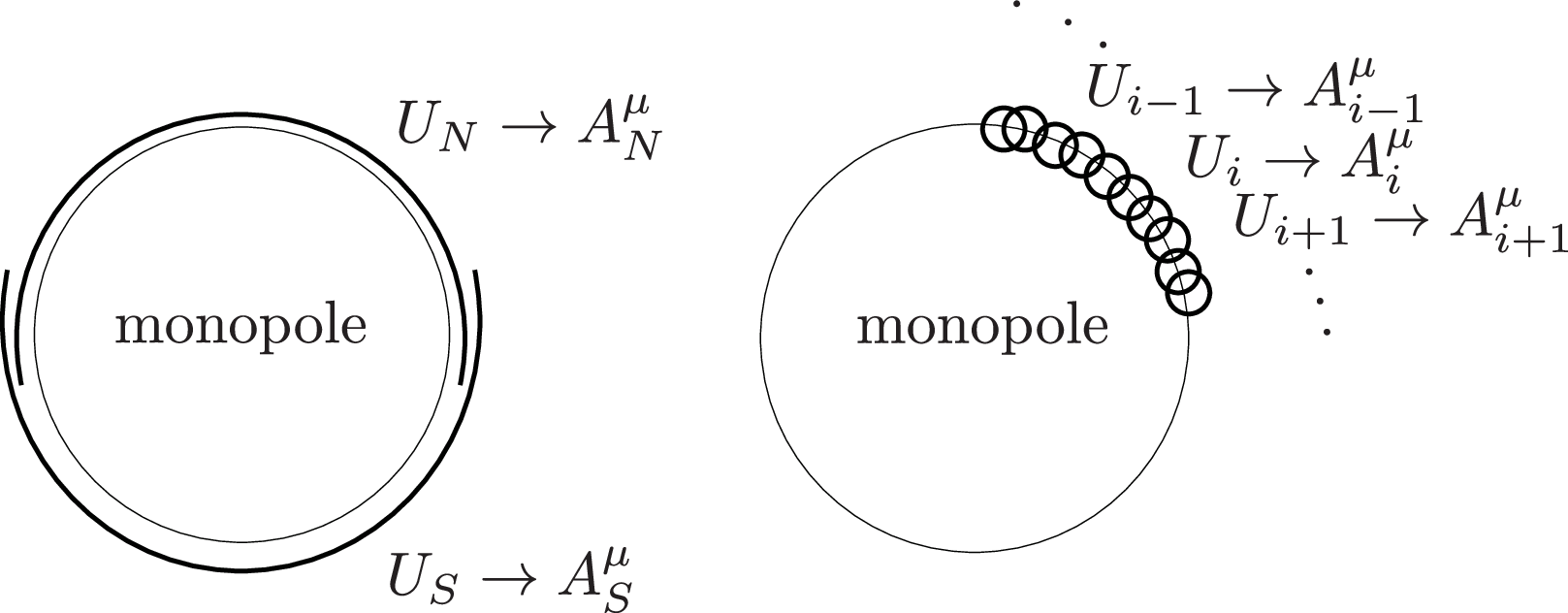}
 \caption{In the left panel, the surface of the sphere is covered by two
 open coverings, and two solutions are constructed in total.
 The two solutions are connected by a gauge transformation.
In the right panel, the sphere is covered by open coverings as per
 the most basic definition of the covariant derivative.
These are called ``bundles''. As local gauge
 transformations are used in defining the covariant derivative, gauge
 transformations are needed to connect the solutions obtained between
 neighboring open coverings.
Normally, no one uses the definition on the right when constructing a
 monopole solution but it becomes important when discussing stationary
 particle production.} 
\label{fig-monopoles}
\end{figure}

Fig.\ref{fig-2} can also have a different explanation.
Using local trivialization, the covariant derivative is defined in the
open covering ($U_i$) around a point ($t=t_i$).
In writing down the field equations, $A^\mu$ was fixed as in
Eq.(\ref{eq-vecpot}), which compromised the degrees of freedom of the
theory and restricted the range of application of the equation
to the vicinity of $t=t_0$.
Therefore, if one wants to define the covariant derivative in the range
$0 < t < 1$, one has to bundle the open coverings $U_i$ defined around $t_i=i/N
,i=0,1,...N$ to have $\lim_{N\rightarrow \infty}\bigcup_{i=0,..N} U_i$.
Here, for each $U_i$ one has to define 
\begin{eqnarray}
A^\mu&=&(0,0,0, -E (t-t_i)).
\end{eqnarray}
This mathematical procedure of the manifold explains stationary particle production
at any time.
In the case of monopoles, only the two $U_N$ and $U_S$ are enough to
describe the solution 
because it is about topology, but in the case of particle generation,
the procedure must be followed according to the
fundamentals.
For comparison, Fig.\ref{fig-monopoles} illustrates how to construct a
``clever'' solution and a ``stick-to-the-basics'' solution for a monopole.
It is by no means a self-evident issue that the left construction method coincides with the right construction method. 

\begin{figure}[t]
\centering
\includegraphics[width=1.0\columnwidth]{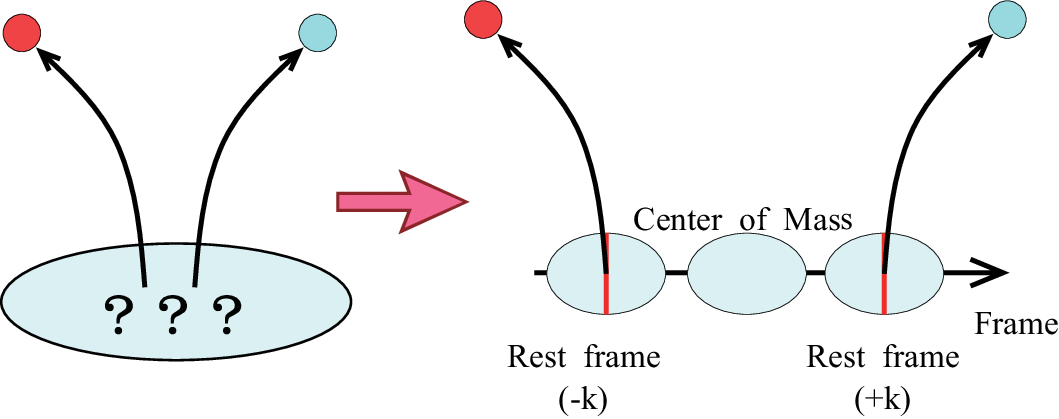}
 \caption{This figure shows the relationship between the Stokes lines and
 the frames when $k_z\ne 0$ in the center of the mass frame. The
 original figure can be found in Ref.\cite{Matsuda:2023mzr}.}
\label{fig-pmk}
\end{figure}

To understand the importance of the definition of the local vacuum
on the manifold, which incorporates the freedom of the gauge and general relativity, 
note that $k_z\ne 0$ shifts the position of the Stokes line from $t=t_0$
(i.e, $A^\mu=0$). 
This shifts the Stokes lines away from the defined ``vacuum'' and spoils
the scenario of stationary radiation.
The choice of the reference frame is crucial, even if the Schwinger effect deals with
gauge theory and not obviously with general relativity.
We show in Fig.\ref{fig-pmk} the displacement of the Stokes line from
the center-of-mass frame and how
it coincides with the defined vacuum in the rest frames.
We omitted explicit Lorentz factors of $E$ and $\omega$.

We have explained stationary particle production of the Schwinger effect
using the field equation with the help of manifolds.
The Stokes phenomena of the Schwinger effect are defined without 
relying on artificial asymptotic states.
The definition of the ``vacuum'' on the manifold is consistent with
stationary radiation. 
The freedom of the gauge and general relativity is thus properly
incorporated into the field equation using the manifold.

We have seen that once the field equation is considered on manifolds,
 one can find the local vacuum of particle creation.
Our definition of the vacuum is a natural extension of the vacuum
 in general relativity to gauge theory.

So far, our calculation is fully consistent with Schwinger's
original calculation.
However, after calculating the local Unruh effect, we will find a
significant difference. 

For the Schwinger effect, the Stokes phenomenon is obvious in the field
equation. 
However, this is not true for Hawking radiation
and the Unruh effect.
The cause of such failure tells us a significant difference between
the local inertial frame and the Lorentz frame.

\section{How to find the Stokes phenomenon for the Unruh effect and Hawking radiation}
We have seen that for the Schwinger effect, one can define a local vacuum
on the manifold to find stationary radiation due to the Stokes phenomenon.
The vacuum of the particle production is defined by choosing a frame for which each
particle is at rest.

Let us think about the definition of the ``vacuum'' on the manifold.
In general relativity, there are two different coordinate
systems for which the ``vacuum'' can be defined: the Lorentz
frame and the local inertial frame.
Although the two frames are often considered to be physically
identical (because they give the same metric), the difference is crucial
for our discussion.

In order to understand the difference from the Schwinger effect, let us
see what happens if the derivatives of the tangent space are replaced by
covariant derivatives. 
This manipulation gives the Klein-Gordon equation of an accelerating
observer (Unruh) or on curved spacetime (Hawking), but unlike the
Schwinger effect, it cannot explain the 
local Stokes phenomenon\cite{Enomoto:2022mti}.\footnote{Stokes
phenomena may occur around black holes.
See Ref.\cite{Enomoto:2022mti,Dumlu:2020wvd} for their contribution to
the glaybody factor.
We discuss Stokes phenomena that are directly related to local particle
production.}
The origin of this problem is the fact that the equations of motion
of the accelerating system are constructed from the Lorenz system
and they are not directly looking at the inertial system. 
To understand the situation, one can calculate the vierbein of the
inertial system in the Rindler coordinate.
Then one will find that it has off-diagonal elements.
To define the covariant derivatives on manifolds, local
trivialization has to be used.
Since the local trivialization gives a direct product space, it
 cannot give the inertial system.
The mathematical definition of the covariant derivatives uses the Lorenz
frame.

Therefore, since the Unruh effect and Hawking radiation are
described for the inertial vacuum, what we have to consider is the
vierbein, not the field equations.
However, there has been no analysis of the Stokes phenomenon using the
vierbein until Ref.\cite{Enomoto:2022mti}.
Of course, even if the field equations are used, the Bogoliubov
transformation can be computed by extrapolating the solutions and
analyzing them in the whole space.
However, such an extrapolation has a risk that extra information may be
introduced by the procedure.
Our speculation is that a strong correlation between distant wedges in the
conventional calculation of the Unruh effect is nothing but the 
``extra information''.
We refer to this problem as the factor 2 problem, from the reason that
will be stated below.

To understand the essence, it would be better to consider the
Unruh effect before Hawking radiation.
In the Unruh effect, the vacuum seen by an accelerating observer is the
inertial system, not the Lorentz system. 
One can compute the vierbein to confirm that the vacuum of the Unruh
effect is defined for the inertial system.
The metric is the same whether it is a local inertial
system or a Lorentz system, so we rarely distinguish between the
two, but in the present case, the difference is crucial.
In fact, the vierbein of the Rindler spacetime can be calculated from
\begin{eqnarray}
\label{eq-rindler-t}
t&=&\frac{1+\alpha x_r}{\alpha}\sinh (\alpha t_r)\nonumber\\
x&=&\frac{1+\alpha x_r}{\alpha}\cosh (\alpha t_r),
\end{eqnarray}
which describes the coordinate system of an object
moving at constant acceleration $\alpha$ through a flat space-time
represented by $(t,x)$.
One will find 
\begin{eqnarray}
dt&=&\left(1+\alpha x_r\right)\cosh (\alpha t_r) dt_r
+\sinh(\alpha t)dx_r\nonumber\\
dx&=&\cosh(\alpha t)dx_r +\left(1+\alpha x_r\right)\sinh (\alpha t_r) dt_r.
\end{eqnarray} 
Obviously, the vierbein has off-diagonal elements and does not support
local trivialization by itself..
On the other hand, the metric is calculated as
\begin{eqnarray}
g_{\mu\nu}&=&\eta_{mn}e^m_\mu e^n_\nu,
\end{eqnarray}
where $\eta_{mn}$ is for the local Minkowski space.
This gives the Rindler metric given by 
\begin{eqnarray}
\label{eq-metric-R}
ds^2&=&-(1+ \alpha x_r)^2dt_r^2+dx_r^2.
\end{eqnarray}
Note that the metric is identical for both (inertial and the Lorentz)
frames.
As we will see, the vierbein of the inertial system is where the Stokes
phenomenon of stationary radiation on curved space-time comes from,
while the field equations are defined for the Lorentz frame.

As is summarized in the textbook\cite{Birrell:1982ix}, one can 
calculate the Bogoliubov coefficients by considering carefully 
the global structure of the Rindler coordinates and the relationship
between the vacuum solutions written in the two coordinate systems.
The question, of course, is why it is not possible to find the Stokes
phenomenon locally, as in the case of the Schwinger effect.

Let us elaborate on the global calculation first to make 
the reason for the distant correlation clear.
We introduce the Rindler coordinates $(\tau,\xi)$ in the right wedge as
\begin{eqnarray}
t&=&\frac{e^{\alpha\xi}}{\alpha}\sinh a\tau\nonumber\\
z&=&\frac{e^{\alpha\xi}}{\alpha}\cosh a\tau,
\end{eqnarray}
where $(t,z)$ are the coordinates of the Minkowski space.
Introduce the light-cone coordinate system of the Rindler space as
$u=\tau-\xi, v=\tau+\xi$.
One can see that they are connected to the
light-cone coordinate system of the Minkowski space $(U,V)$ as
\begin{eqnarray}
U&=&-\frac{e^{-\alpha u}}{\alpha}\nonumber\\
V&=&\frac{e^{\alpha v}}{\alpha}.
\end{eqnarray}
We define the Rindler coordinates $(\tilde{\tau},\tilde{\xi})$
in the left wedge.

We consider massless particles for simplicity and consider only the
right-moving waves.
In the inertial system, the field $\phi(U)$ is expanded as
\begin{eqnarray}
\phi(U)&=&\int^\infty_0dk \left[a_k f_k^{(M)}(U)+a_k^\dagger
			   f_k^{(M)*}(U)\right],\nonumber\\
f_k^{(M)}(U)&=&\frac{e^{-ikU}}{\sqrt{4\pi k}},
\end{eqnarray}
which defines the inertial vacuum by $a_k|0_M\rangle=0$

For the right Rindler system $(\tau,\xi)$, we have
\begin{eqnarray}
\phi_R(u)&=&\int^\infty_0dp \left[b_p^{(R)} f_p^{(R)}(u)+b_p^{(R)\dagger}
			   f_p^{(R)*}(u)\right],\nonumber\\
f_p^{(R)}(u)&=&\frac{e^{-ipu}}{\sqrt{4\pi p}}=\theta(-U)\frac{(-\alpha
 U)^{ip/\alpha}}{\sqrt{4\pi p}},
\end{eqnarray}
which defines the Rindler (right) vacuum by $b_p^{(R)}|0_R\rangle=0$,
and same calculation defines the Rindler (left) vacuum.
Using the above solutions one can find\cite{Iso-textbook}
\begin{eqnarray}
\label{eq-def-vac-unruh}
|0_M\rangle&\propto&\exp\left[-\prod_p 
\left(e^{-\pi p/2\alpha}b_p^{L\dagger}\right)
\left(e^{-\pi p/2\alpha}b_p^{R\dagger}\right)\right]|0_L\rangle\otimes|0_R\rangle,
\end{eqnarray}
which shows a strong correlation between distant wedges.
After taking normalization and trace about the left Rindler states, one will
find that the Unruh temperature is $T_U=\hbar \alpha/2\pi c k_B$. 
What is important is the duplication of the factor 
$e^{-\pi p/2\alpha}$ due to the correlation.
This is the source of our factor 2 problem and of course, this factor
does not appear in local calculation.

One might insist that the global calculation ``revealed'' a surprising
correlation between two causally disconnected wedges.
However, it is quite unnatural that global information is essential for
the calculation when
its motion can be viewed as constant acceleration only over a certain
period of time.

Now let us see how the Stokes phenomenon of the Unruh effect appears.
First remember that during the Unruh effect, an accelerating observer is looking at the
inertial vacuum.
In that case,  the vacuum solutions of the inertial system have to be seen by
an observer using the vierbein.
We use $dt=\cosh (\alpha t_r) dt_r$
to write the vacuum solutions into
\begin{eqnarray}
\label{eq-rindler-sol}
\phi_k^\pm(t)&=&A_k e^{\pm i \int \omega dt}\nonumber\\
&=&A_k e^{\pm i \int \omega_k \cosh(\alpha t_r) dt_r}.
\end{eqnarray}
Note that we are choosing the particle's rest frame.
Normally, it is quite difficult to recognize the Stokes phenomenon of these
solutions.
However, using the exact WKB\cite{Enomoto:2020xlf, Enomoto:2021hfv,
Virtual:2015HKT}, one can solve the problem.
First, define $Q(t)_0\equiv -\omega_k^2\cosh^2(\alpha t_r)$ and consider
 the ``Schr\"odinger'' equation
\begin{eqnarray}
\label{eq-rindler-EoM}
\left(-\frac{d^2}{dt^2}+\eta^2 Q(t,\eta)\right)\psi(t,\eta)&=&0,
\end{eqnarray}
where $\eta\gg 1$. $Q(t,\eta)$ is expanded as
\begin{eqnarray}
Q(t,\eta)&=&Q_0(t)+\eta^{-1}Q_1(t)+\eta^{-2}Q(t)+\cdots.
\end{eqnarray}
Note that $\hbar\ll 1$ of quantum mechanics has been replaced by
$\eta\gg 1$ according to mathematical convention.
It is important to note that $\eta$ is not just a large parameter, but a
unique parameter that governs singular perturbations.
Moreover, since $\eta$ is analytically continued to the complex $\eta$-plane, the
concept of the exact WKB is rather different from the conventional WKB approximation with small
$\hbar$.

The solution of this equation can be written as $\psi(t,\eta)\equiv
e^{\int S(t,\eta) dt}$.
Here, $S(t,\eta)$ can be expanded as 
\begin{eqnarray}
S&=&S_{-1}(t)\eta +S_0(t)+S_1(t)\eta^{-1}+\cdots.
\end{eqnarray}
The point of this argument is that after introducing $\eta$ properly
in Eq.(\ref{eq-rindler-sol}), one can choose $Q_i(t), i\ge 1$ to
find the ``Schr\"odinger equation'' of the exact WKB.

\begin{figure}[t]
\centering
\includegraphics[width=0.8\columnwidth]{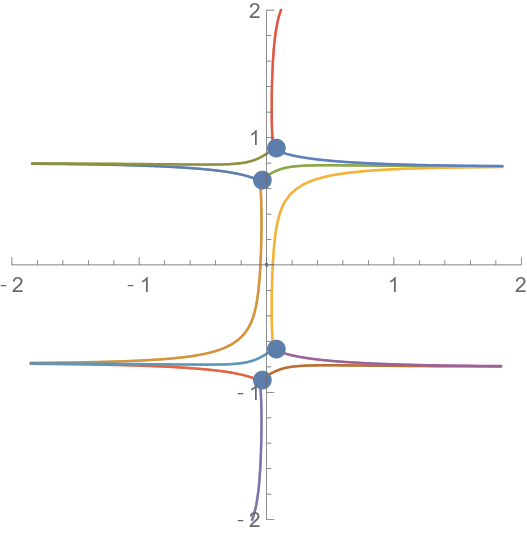}
 \caption{The Stokes lines of the Unruh effect are shown for
 $Q(t)=-(\cosh^2 (2t)-0.05+0.05 i)(1+0.05i)$. The degenerated Stokes
 lines are separated introducing small parameters. The original figure
 can be found in Ref.\cite{Matsuda:2023mzr}.}
\label{fig-Unruh-stokes}
\end{figure}

This procedure allows one to make use of a powerful analysis of the
exact WKB.
One can calculate the Stokes lines only by using 
$Q_0(t)$.\footnote{Remember that $\eta$ is the unique parameter that
governs the singular perturbation. The reason why $Q_0(t)$ is special
is explained in Ref.\cite{Virtual:2015HKT} and the references therein in
the light of singular perturbation theory.}
After drawing the Stokes lines, one can understand that the Stokes line
crosses on the real axis at the origin\cite{Enomoto:2022mti}.
The Stokes lines of the Unruh effect are shown in
Fig.\ref{fig-Unruh-stokes}.
This allowed us to expand $Q(t)_0$ near the origin and finally, we have
\begin{eqnarray}
Q(t)_0&=& -\omega_k^2\cosh(\alpha t_r)\nonumber\\
&\simeq&-\omega_k^2-\alpha^2 \omega_k^2 t_r^2,
\end{eqnarray}
which gives the typical Schr\"odinger equation of scattering by an
inverted quadratic potential\cite{Enomoto:2022mti, Enomoto:2020xlf,Enomoto:2021hfv}.
Again, as in the case of the Schwinger effect, the Stokes lines coincide
with the vacuum only when the vacuum is defined for the rest frame of
the particles.
This frame is not the experimenter's frame.
This choice of the frame is of course consistent with 
experimental analyses\cite{DiPiazza:2011tq}.

The above calculation suggests the Boltzmann factor $\sim e^{-\pi
\omega_k/\alpha}$. 
On the other hand, $\sim e^{-2\pi \omega_k/\alpha}$ was calculated in
the global(conventional) calculation.
They are showing the factor 2 problem.
As we have explained, duplication of the factor due to the
entanglement between disconnected wedges is the origin of this problem.
Whether this correlation actually exists will be
verified by experimentation, since the crucial
evidence appears in the Unruh temperature.

Our calculation also applies to Hawking radiation.
Hawking radiation requires creation of a pair of negative and positive energy
particles across the horizon.
Then, only the positive energy particle outside the horizon can be observed as radiation.
This situation is typical ``2 for 1''.
Namely, two particles are produced but only one is observable.
In this case, the production probability of one particle ($P_1$) must
be discriminated from the observation probability of one particle
($P_1^{obs}$). 
To be more precise, our local analysis distinguishes ``one particle production
rate $P_1$'' from ``one particle observation rate $P_1^{obs}(=$
two particle production rate $(P_1)^2$)''. 
Considering these discrepancies, one can find that there is no factor 2
problem in Hawking radiation because the ``2 for 
1'' particle production is essential in Hawking radiation.

\section{How do the Unruh effect and the Schwinger effect coexist?}
Finally, it should be mentioned that the Unruh effect\footnote{This
corresponds to Hawking radiation without 
the event horizon, because the strong electric field allows pair
creation. The crucial difference is that both particles are observed.} 
occur simultaneously with the Schwinger effect if they are both defined
on the manifold.
We will focus on the differences between them in
order to avoid a superficial discussion.

The acceleration of a charged particle in a strong electromagnetic field
is $\alpha=qE/m$, which, when used in ``our'' equation for
the Unruh effect, gives the same coefficient as
the Schwinger effect.

We have seen that the source of the Schwinger effect is the vector
potential and it can be seen by the field equation on the manifold.
On the other hand, the Unruh effect is explained by the vierbein of the
inertial system, not directly by the field equation.

As noted above, the way to choose the vacuum in the Unruh effect is to
 ``look at the local inertial system, choosing rest frame of 
 the particle (to be produced)''. 
Considering that the  generated particles are
accelerated by the electric field, the particles are generated from 
``the vacuum seen by an accelerating observer (particle)''.
This is precisely the same situation as in the case of the Unruh effect.
One thing that is still not clear is whether the vacuum defined for the
 Schwinger effect is identical to the inertial vacuum in the Unruh effect.

Unlike the Unruh effect, pair production is possible
without the event horizon, if the electric field is strong enough.
In this case, since one of the two particles of a pair 
does not disappear into the horizon, both particles are observed.
This is not a ``2 for 1'' particle production and
the Unruh temperature observed simultaneously with the
Schwinger effect should show a factor of 2 difference compared to
 Hawking radiation.

Perhaps the most interesting aspect of this story is whether it is
possible to verify experimentally that these effects
occur simultaneously.
As we have discussed above, the two effects arise from separate physical
phenomena.
The differences between the two are listed below.
\begin{itemize}
\item The Stokes phenomena of the Schwinger effect appears from the
      field equation, while the Stokes phenomena of the Unruh effect
      appears from the vierbein.
\item Since the Schwinger effect appears in the field equation, the vacuum of
the Schwinger effect would be the Lorentz frame, while the Unruh effect
looks at the local inertial frame.
\end{itemize}
Therefore, it is still not clear whether they could contribute in the
same way or not.
What is important for the discussion is the definition of the vacuum on
the manifold.
If the vacuum were defined as an asymptotic state, such an argument
would never have arisen.

Let us think about possibilities.
When written in the exact WKB notation, there might be 
$Q_0^{Schwinger}(t)$ and $Q_0^{Unruh}(t)$ to
give $Q^{total}_0(t)=Q_0^{Schwinger}(t)+Q_0^{Unruh}(t)$.
In this case, the coefficient of the inverted quadratic potential
has to be the sum of the two contributions. 
On the other hand, it is not surprising that the
Schwinger and Unruh effects are defined for different vacuum states and
they can occur as independent phenomena. 
In that case, the result should be
$\beta^{tot}=\beta^{Unruh}+\beta^{Schwinger}$.
At this moment we cannot determine which is correct, and it  has to be
determined only by experimental verification.

If our considerations are correct, the experimental results must be
significantly different from Schwinger's calculation.
The previously mentioned problem with factor 2 of the Unruh effect 
can also be confirmed by experiment.
If, on the other hand, the amplification noted here did not occur, then
it can be concluded that the Schwinger and Unruh effects are in fact
identical physical phenomena that are linked even more deeply in the
theory.

\section{Concluding remarks}
I first read Hawking's paper on Hawking radiation when I was in a
master's program. 
At the time, I had just moved from pharmaceutical sciences to a
postgraduate degree in physics, so I didn't know what was right or
wrong. 
Even so, I couldn't stand the unnaturalness of Hawking's calculations.
His calculation was to prepare a vacuum in the infinitely distant
infinite past and future, and connect the two via the collapse of a
black hole. 
As a physical phenomenon, it deals with the production of particles on
the surface of a black hole. 
I thought it was strange if calculations could not be done in the
vicinity of what is supposed to be a black hole. 
For me, his calculation looked like a painstaking work of a genius with
superhuman arithmetic skills. 
As a newcomer from another field, I thought there was something deeper
that I had not yet been able to recognize. 
So I investigated how similar physical phenomena were handled in physics.
I first looked at the Unruh effect. 
What I understood was that the Unruh effect also requires the same
extrapolation as the Hawking radiation, and that there seems to be
entanglement at a distance in the Rindler space. 
Having wondered about these situations, after many years, I looked into
the Schwinger effect. 
While studying the Schwinger effect, I came across the calculation that
I had hoped for.
It was the Stokes phenomenon from the field equation. 
But it still seemed incomplete to me.
In the discussion of the field equations, the vacuum is still defined as a
distant asymptotic state, and the unnaturalness I felt with Hawking
radiation was present here too. 
Schwinger's calculation, which claims to have calculated the decay
rate of the vacuum, did not involve a ``real'' vacuum transitioning into
another vacuum, in the sense that no domain walls are formulated around
the new vacuum.
It looked to me like there was some kind of deception going on.

Then I came across a way of constructing a monopole solution. 
When the monopole solution is constructed from the equation of motion,
the two solutions are pasted together and connected by a gauge
transformation between them. 
I realized that the reason for this cumbersome procedure is that
the original mathematical manifold is not viewed as it is in the field
equation.
I also thought that the vacuum must be properly defined on the manifold
that represents the whole theory. 

The reason why similar calculations (the Stokes phenomena) could not be
done with the accelerating system was also a mystery. 
After noticing a clue, further years were needed before the Stokes
phenomenon was discovered in the accelerating system. 
It is unfortunate for students that no textbook mentions why in the
Unruh effect (Hawking radiation) naive use of field equations cannot
explain particle production, since many students have to learn
unnatural extrapolations and complicated calculations without knowing
why such calculations are necessary. 

We believe we have solved these problems with our study. 
This research began with a simple question from my student days and
hopefully, this research will help students struggling with the
same questions.

\section{Acknowledgments}
The author would like to thank all the participants and the organizers
of the conference, and he is deeply grateful to Professor Vesselin
Petkov.
The author was particularly encouraged by the comments given by
Professor Orfeu Bertolami.

\end{document}